\documentclass{webofc}

\usepackage[varg]{txfonts}   
\usepackage{hyperref}
\usepackage{url}
\hypersetup{colorlinks=true,citecolor=blue,urlcolor=blue,linkcolor=blue}

\usepackage{slashed}

\newcommand{\avg}[1]{\langle #1 \rangle}
\newcommand{\fa}{f_a}
\newcommand{\etp}{\eta'}
\newcommand{\Mpi}{M_\pi}
\newcommand{\MK}{M_K}
\newcommand{\gaP}{\gamma_{aP}}

\begin{document}

\title{Phenomenology of axion--meson mixing and scattering processes}

\author{%
  \firstname{No\'e} \lastname{Duarte-Gonz\'alez}\inst{1,2}\fnsep\thanks{\email{nduarte@ifae.es}}
  \fnsep\thanks{
  I would like to express my gratitude to the MESON2026 Organizing Committee for the opportunity of presenting this contribution.
  This work is supported by the Ministerio de Ciencia, Innovación y Universidades under Grant No.~PID2023–146142NB-I00 and through the FPU programme (FPU23/03545).}
}

\institute{%
  Grup de F\'isica Te\`orica, Departament de F\'isica, Universitat Aut\`onoma de Barcelona, 08193 Bellaterra (Barcelona), Spain
  \and
  Institut de F\'isica d'Altes Energies (IFAE) and The Barcelona Institute of Science and Technology (BIST), Campus UAB, 08193 Bellaterra (Barcelona), Spain
}

\abstract{%
We study the mixing between a generic axion-like particle (ALP) and the light pseudoscalar mesons within chiral perturbation theory ($\chi$PT), and its consequences for low-energy scattering and decay processes. Starting from the most general ALP effective Lagrangian, we derive the ALP-meson kinetic and mass mixing induced at leading order in the ALP decay constant and isospin breaking for arbitrary ALP masses. As a result, we obtain the physical ALP mass to $\mathcal{O}(F^2/\fa^2)$, recovering the QCD-axion limit and finding that any ALP with a vanishing bare mass shares the QCD-axion mass. We provide a compact parameterization of every four-body amplitude involving pseudoscalar mesons and an arbitrary number of ALPs. As an illustration, we show the $K\bar{K}\to P^0a$ scattering. The framework applies for general ALP masses and couplings and consistently incorporates isospin breaking, essential for otherwise forbidden processes. This contribution summarizes the main results of \cite{InPreparation}.
}

\maketitle

\section{Introduction}
\label{sec:intro}
The strong $\mathrm{CP}$ problem \cite{Axions2008} remains one of the sharpest open questions of the Standard Model, and its most compelling solution predicts a light pseudoscalar, the QCD axion~\cite{PecceiQuinn,PecceiQuinn2,Weinberg,Wilczek}. More generally, many extensions of the Standard Model contain light pseudoscalar bosons, dubbed as axion-like particles (ALPs), whose masses and couplings are not tied to a single relation. ALPs thus provide a broad and well-motivated portal to new physics, and their interactions with the lightest pseudoscalar mesons are naturally described within chiral perturbation theory ($\chi$PT)~\cite{GeorgiKaplanRandall}.

Most $\chi$PT studies of axion--meson dynamics have been restricted either to the QCD axion or to the regime of very light masses, $m_a\ll\Lambda_{\rm QCD}$ \cite{Notari}. Mass effects, however, can be phenomenologically relevant, in cosmology and in meson decays, and there is growing interest at meson factories in probing ALPs directly in meson decays, including the heavier $\etp$ \cite{GanKubis}. In addition, isospin-breaking (IB) effects are frequently neglected. While this is justified for isospin-allowed reactions, where IB represents a small correction, it is essential when the process is otherwise forbidden.

In this contribution, we investigate four-body processes ($2\to2$ scattering and $1\to3$ decays) involving ALPs and pseudoscalar mesons within $\chi$PT, including linear IB effects. The results are meant as a first, general survey and as the starting point of a dedicated effort incorporating higher-order and rescattering effects; full details will be reported elsewhere.

\section{Formalism and conventions}
\label{sec:formalism}
We start from the most general effective Lagrangian describing an ALP, $a$, coupled to Standard Model particles with masses at or below $1$~GeV: light quarks $q=u,d,s$, gluons, light charged leptons $\ell=e,\mu$ and photons~\cite{GeorgiKaplanRandall,GanKubis}:
\begin{align}
\mathcal{L}_{\rm ALP} &= 
-\frac14G_{\mu\nu}^aG^{a,\mu\nu}+
\bar{q}(i\slashed{D}-\mathcal{M}_q)q+
\bar\theta\dfrac{\alpha_s}{8\pi}G^a_{\mu\nu}\tilde{G}^{a,\mu\nu}
+\frac12(\partial_\mu a)(\partial^\mu a)-\frac12 m_0^2 a^2
-\nonumber\\ &
-\frac{\alpha_s}{8\pi}\frac{a}{\fa}G^a_{\mu\nu}\tilde G^{a,\mu\nu}
-c_q\frac{\partial_\mu a}{2\fa}\bar q\gamma^\mu\gamma_5 q-c_\gamma\frac{\alpha}{8\pi}\frac{a}{\fa}F_{\mu\nu}\tilde F^{\mu\nu}
-c_\ell\frac{\partial_\mu a}{2\fa}\bar\ell \gamma^\mu\gamma_5 \ell\,,
\label{eq:LALP}
\end{align}
with $\tilde G^{a,\mu\nu}=\tfrac12\epsilon^{\mu\nu\alpha\beta}G^a_{\alpha\beta}$ (we take $\epsilon^{0123}=+1$) and $\mathcal{M}={\rm diag}(m_u,m_d,m_s)$ the quark-mass matrix. We also assume flavor and $\mathrm{CP}$-conserving couplings to quarks and leptons through axial currents\footnote{We will also write the ALP-quark couplings as a diagonal matrix in flavor space, $c={\rm diag}(c_u,c_d,c_s)$, which is a more suitable notation for the hadronic theory.}.

Due to the chiral anomaly, the ALP-gluon coupling can be removed by a chiral rotation of the quark fields\footnote{We could treat $q=(u,d,s)$ as a vector containing the light quarks, and $\kappa$ being a matrix in flavor space that characterizes the chiral rotation. However, since we do not consider flavor-violating couplings, $\kappa$ is a diagonal matrix and we write the rotation for each individual quark.},
$
q\to \exp\left(i
a
\kappa_q \gamma_5
/f_a\right)q
$,
with $\avg{\kappa}=\kappa_u+\kappa_d+\kappa_s=1$. The convenient choice is $\kappa=\mathcal{M}^{-1}/\avg{\mathcal{M}^{-1}}$, which will eliminate the ALP-octet mass mixing.

After applying the chiral rotation to Eq.~\eqref{eq:LALP}, the ALP-gluon term reappears as a shift in the ALP-quark axial coupling ($c\to\bar{c}=c+\kappa$) and ALP-photon coupling ($c_\gamma\to\bar{c}_\gamma=~c_\gamma-2N_c\avg{\kappa Q^2}$, with $Q$ the quark charge matrix). Additionally, the quark mass matrix is modified and acquires a non-linear ALP dependence, $\mathcal{M}\to\hat{\mathcal{M}}(a)=\exp\left(ia\kappa\gamma_5\right/\fa)\mathcal{M}$.

At the hadronic level, the theory is described by the leading-order (LO), large-$N_c$ chiral Lagrangian~\cite{KaiserLeutwyler,GasserLeutwyler,SchererSchindler}. In the presence of the ALP as an external field described by Eq. \ref{eq:LALP}, the chiral Lagrangian takes the form\footnote{We will focus most of the work on ALP-meson interactions, and will omit the leptons and photons from the effective description.}
\begin{align}
\mathcal{L}_{\rm eff}&
=\dfrac{F^2}{4}\langle(\partial_\mu U)(\partial^\mu U^\dagger)\rangle+
\dfrac12(\partial_\mu a)(\partial^\mu a)+
\frac{F^2}{4}\langle\hat{\chi}(a) U^\dagger+\text{h.c.}\rangle-
\dfrac12M_0^2\eta_0^2-
\dfrac12m_0^2a^2-
\nonumber\\&-
i\dfrac{F^2}{4f_a}(\partial_\mu a)\left\langle\bar{c}\left(U^\dagger\partial^\mu U-U\partial^\mu U^\dagger\right)\right\rangle+
\dfrac{F^2}{8f_a^2}(\partial_\mu a)(\partial^\mu a)\left\langle\bar{c}\left(U^\dagger\bar{c}U+\bar{c}\right)\right\rangle
\,,
\label{eq:Leff}
\end{align}
where $U=\exp\left(i\sqrt2\Phi/F\right)$ collects the pseudoscalar nonet in $\Phi$, $F\simeq F_\pi\simeq 92.4$~MeV is the LO decay constant, $M_0$ parametrizes the singlet mass in the chiral limit and the scalar source operator, $\chi$, is modified by the presence of the ALP in the quark mass matrix, $\hat{\chi}(a) = \exp\left(-ia\kappa/f_a\right)\chi$.

\section{ALP--meson mixing and physical states}
\label{sec:mixing}
Expanding Eq.~\eqref{eq:Leff} to quadratic order in the neutral fields $\phi^T=(a,\pi_3,\eta_8,\eta_0)$ generates non-canonical kinetic terms, inducing kinetic mixing, and mass-mixing terms,
\begin{equation}
\mathcal{L}\supset\frac12(\partial_\mu\phi)^T K(\partial^\mu\phi)
-\frac12\phi^T \mathcal{M}_{\rm mix}^2\phi.
\label{eq:mix}
\end{equation}
The physical fields are obtained from the mathematical ones following a two-step process, making the kinetic term canonical and then diagonalizing the mass matrix. That way, we get $\phi = \sqrt{Z}\hat{\phi}=\sqrt{Z}R^T\phi_{\rm phys}$, where $\sqrt{Z}^{\,2}=K^{-1}$ and $R$ is the orthogonal matrix that diagonalizes the mass term after the kinetic term is canonical:
\begin{align}
\mathcal{L}\supset
\dfrac12(\partial_\mu\hat\phi)^T(\partial^\mu\hat\phi)-
\dfrac12\hat\phi\sqrt{Z}M_{\rm mix}^2\sqrt{Z}\hat{\phi}\,\,\Rightarrow\,\,
\hat{M}^2 = \sqrt{Z}M_{\rm mix}^2\sqrt{Z} = R^TM^2_DR\,,
\label{eq:diagonalization}
\end{align}
with $M_D^2$ diagonal containing the physical masses. Given that $R$ is a dimension 4 orthogonal matrix, it can be decomposed as a product of 6 different rotations:
\begin{align}
R = R_{a\etp}
R_{a\eta}
R_{a\pi}
R_{\pi\etp}
R_{\pi\eta}
R_{\eta\etp}\,,
\label{eq:rotation}
\end{align}
where each $R_{PP'}$ is a function of a single angle, $\theta_{PP'}$, and it represents a rotation affecting only the pseudoscalars $P$ and $P'$.

Out of the 6 angles that appear in Eq.~\eqref{eq:rotation}, all but one ($\theta_{\eta\eta'}$) are assumed to be small: the $\pi$-$\eta^{(\prime)}$ are known to be proportional to an IB parameter of order $1\%$, and the $a$-$P$ angles are assumed to be proportional to the very small ratio $F/f_a$. It is worth pointing out that, at linear order in $F/f_a$, the physical ALP field appear as a shift in the mathematical meson fields, $P\to P+\gamma_{aP}a$, with $\gamma_{aP}$ some effective angle. Full expressions of these effective angles, including IB corrections to the ALP angles, are included in \cite{InPreparation}.

\section{The physical ALP mass}
\label{sec:mass}
The physical ALP mass is obtained by diagonalizing $\hat{\mathcal{M}}^2$ from Eq.~\eqref{eq:diagonalization} to $\mathcal{O}(F^2/\fa^2)$; no term of $\mathcal{O}(F/\fa)$ contributes\footnote{This implies, in particular, that $m_a=m_0$ in LO amplitude calculations.}, since a correction to the ALP propagator requires at least two insertions of the ALP--meson mixing operators, each weighted by $F/\fa$. Starting from the QCD axion, recovered for $m_0=0$ and $c_u=c_d=c_s=0$, diagonalization yields
\begin{equation}
m^2_{a,\rm QCD}
=\left(\frac{F}{f_a}\right)^2\frac{M_0^2M_\pi^2(2\MK^2-\Mpi^2)}{6M_\pi^2(2M_K^2-M_\pi^2)+2M_0^2(4M_K^2-M_\pi^2)}
\Rightarrow
m_{a,\rm QCD}
=6.16~\text{eV}\,\dfrac{10^6~\text{GeV}}{f_a}\,.
\label{eq:mQCD}
\end{equation}
In the the $M_0,m_s\to\infty$ limits (integrating out the $\etp$, $\eta$ and kaons), Eq.~\eqref{eq:mQCD} reduces to the well-known two flavor value of the QCD axion mass \cite{Axions2008}, \begin{align*}
m_{a,\rm QCD}\big|_{\text{SU}(2)}=\frac{FM_\pi}{2f_a}=6.34~\text{eV}\,\frac{10^6~{\text{GeV}}}{f_a}\,.
\end{align*}
The two-flavor approximation thus differs from the full result by only about $3\%$.

For general $m_0$ and quark couplings, the mass reads
\begin{equation}
m_a^2=m_0^2+m^2_{a,\rm QCD}+f^{(0)}+\sum_q f^{(q)}c_q+\sum_{q,q'}f^{(qq')}c_q c_{q'} ,
\label{eq:mgen}
\end{equation}
where the $f^{(x)}$ are rational functions of $m_0^2$. Two properties stand out. First, only the $m_0^2$ term is unsuppressed: all remaining contributions are of $\mathcal{O}(F^2/\fa^2)$. Second, every $f^{(x)}$ is proportional to $m_0^2$ and therefore vanishes at $m_0=0$, so any ALP without a bare mass has exactly the QCD axion mass, regardless of its axial quark couplings.

We may also parametrize $m_a^2$ by means of the IB powers, $m_a^2=m_0^2+(F/\fa)^2(g_0+\varepsilon g_1)$, with $\varepsilon$ an IB breaking parameter and $g_i$ rational functions of $c_q$ and $m_0^2$. 
The only relevant source of IB breaking is the kinetic $a$-$\pi^0$ mixing, which does not contribute for $m_0=0$. Therefore, $g_1=0$ for $m_0=0$ and only massive ALPs receive linear IB corrections to their mass.

Performing a numerical comparison of $f^{(x)}$ versus $m_{a,\text{QCD}}^2$, we find that most contributions to the ALP mass are highly suppressed for sensible values of $m_0$ and the couplings. More precisely, the ratios $R^{(x)}=f^{(x)}/m_{a,\text{QCD}}^2$ are of order
\begin{align*}
R^{(x)}& \sim10^{-4}\,\dfrac{m_0^2}{1~{\text{MeV}}^2},\quad
x=0,u,d,uu,dd,ss,\\
R^{(s)}&\sim10^{-5}\,\dfrac{m_0^2}{1~{\text{MeV}}^2},
\quad
R^{(ud)}\sim10^{-9}\,\dfrac{m_0^2}{1~{\text{MeV}}^2},\quad
R^{(us),(ds)}\sim10^{-10}\,\dfrac{m_0^4}{1~\text{MeV}^4}.
\end{align*}
Finally, it is worth pointing out that, for any $m_0<M_\pi$, $f^{(x)}<0$ in every case but the quadratic cross-products, $x=ud,us,ds$, meaning that positive (negative) ALP-quark couplings reduce (increase) the ALP mass for sensible values of $c_q$.

\section{Parameterization of four-body amplitudes}
\label{sec:amp}
The diagonalized interaction Lagrangian allows any four-body amplitude ($2\to2$ scattering or a three-body decay) with an arbitrary number of ALPs to be easily parameterized. We write the amplitudes in terms of coefficients named after their parent operator: $c_{s_n}$ (mass-like), $c^{M_1M_2}_{D}$ (derivative), $c^{M}_{A}$ (single axial current) and $c_{AA}$ (double axial current). Superscripts $P,P',\dots\,(=\pi^0,\eta,\etp)$ label the effective mixing angle $\gaP$ through which an ALP is produced. Every ALP operator or mixing angle carries a factor $F/\fa$, so an amplitude with $n$ ALPs scales as $\mathcal{A}=\mathcal{O}((F/\fa)^n)$. Since $\mathcal{L}^{(4)}_{D}$ is built entirely out of commutators, processes involving only ALPs and neutral mesons ($\pi^0,\eta,\etp$) receive no derivative contribution and thus have no momentum depedence.

As a representative example, with $s=(p_{K}+p_{\bar{K}})^2$, the one-ALP amplitude for $K\bar{K}\to P^0a$, with $P^0=\pi^0,\eta,\etp$, reads
\begin{align}
\mathcal{A} & = c_{s_4}^P\gamma_{aP}+c_{s_3}+
\bigg[\left(c_\text{A}^{aK}+c_{\text{D}}^{aK;P}\gamma_{aP}\right)
(s-M_{P^0}^2+m_a^2)-
c_{\text{D}}^{KK;P}\gamma_{aP}
(s-2M_K^2)-\nonumber\\[5pt]
& -\left(c_\text{A}^{aP^0}+c_{\text{D}}^{aP^0;P}\gamma_{aP}\right)
(s-M_{P^0}^2-m_a^2)+
c_{\text{D}}^{KP^0;P}\gamma_{aP}
(s+M_{P^0}^2-m_a^2)
\bigg]\dfrac{1}{2Ff_a},
\label{eq:amp_KK_piax}
\end{align}
with every of the coefficients being defined and compiled in \cite{InPreparation}. Similar parameterizations can be found for $\pi\pi\to\pi a$ scatterings and $\eta^{(\prime)}$ decays, together with discussions about their phenomenology and asymptotic behaviors and matching to existing results \cite{Notari,DiLuzio}.

\section{Conclusions and outlook}
\label{sec:concl}
We have developed a general framework for the mixing of a generic ALP with the light pseudoscalar mesons in $\chi$PT and its low-energy phenomenology. The diagonalization is performed for an arbitrary ALP mass and couplings to light quarks, and includes leading isospin-breaking effects. Our study provides a comprehensive procedure for any four-body amplitude with an arbitrary number of ALPs. As an application, we presented $K\bar{K}\to P^0a$ scattering, but any process can be described. We have further obtained the physical ALP mass to $\mathcal{O}(F^2/\fa^2)$: the bare-mass term is the only unsuppressed contribution, and any ALP with $m_0=0$ shares the QCD-axion mass irrespective of its quark couplings.

Natural extensions include going beyond LO to include higher-order and rescattering effects, systematically adding the ALP-lepton and ALP-photon interactions presented in Sect.~\ref{sec:formalism}, and using the same formalism to parameterize ALP production in the early universe in connection with dark-matter searches.



\end{document}